# Efficient Probabilistic Computing with Stochastic Perovskite Nickelates

Tae Joon Park[1,#,*], Kemal Selcuk[2,#], Hai-Tian Zhang[1,#,*,†], Sukriti Manna[3], Rohit Batra[3], Qi Wang[1], Haoming Yu[1], Subramanian K.R.S. Sankaranarayanan[3,4], Hua Zhou[5], Kerem Y. Camsari[2,*], and Shriram Ramanathan[1,*]

[1]School of Materials Engineering, Purdue University, West Lafayette, IN 47907, USA

[2]Department of Electrical and Computer Engineering, University of California, Santa Barbara, Santa Barbara, CA, 93106, USA

[3]Center for nanoscale materials, Argonne National Laboratory, Argonne, IL 60439, USA

[4]Department of Mechanical and Industrial Engineering, University of Illinois, Chicago, IL 60607, USA

[5] X-ray Science Division, Advanced Photon Source, Argonne National Laboratory, Lemont, IL 60439, USA

[#]These authors contributed equally to this work
[*]Electronic mail: park1080@purdue.edu, htzhang@purdue.edu, camsari@ucsb.edu, and shriram@purdue.edu
[†]Present address: School of Materials Science and Engineering, Beihang University, Beijing 100191, China

## Abstract

**Probabilistic computing has emerged as a viable approach to solve hard optimization problems. Devices with inherent stochasticity can greatly simplify their implementation in electronic hardware. Here, we demonstrate intrinsic stochastic resistance switching controlled via electric fields in perovskite nickelates doped with hydrogen. The ability of hydrogen ions to reside in various metastable configurations in the lattice leads to a distribution of transport gaps. With experimentally characterized p-bits, a shared-synapse p-bit architecture demonstrates highly-parallelized and energy-efficient solutions to optimization problems such as integer factorization and Boolean-satisfiability. The results introduce perovskite nickelates as scalable potential candidates for probabilistic computing and showcase the potential of light-element dopants in next-generation correlated semiconductors.**

# I. Introduction

Probabilistic computing with p-bits has recently emerged as a domain-specific computational paradigm which can accelerate randomized algorithms in many areas related to machine learning and artificial intelligence[1,2]. Specifically, p-bits and p-circuits can be applied to statistical machine learning models such as energy-based neural networks[3], efficient sampling and inference for Bayesian networks[4]. P-circuits have also been demonstrated to accelerate hard optimization problems[5–7] and quantum Monte Carlo algorithms[8,9]. Probabilistic bits have previously been implemented in existing CMOS technology. However these implementations suffer from costly (pseudo) random number generators requiring thousands of transistors[10–12] and poor quality of randomness, increasing energy and area costs for stochastic circuits. Even when the most advanced technology nodes are used, the ultimate scalability of standard digital CMOS-based p-bits are limited. Therefore, nanodevices whose intrinsic randomness can be exploited to design energy-efficient p-bits is highly desirable and represents an early-stage research field. Along these lines, there have been proposals to build probabilistic bits using magnetic and spintronic implementations[13,14] and thermal phase transitions[15,16]. In addition to sources of randomness extracted from nanodevices (such as stochastic magnetic tunnel junctions), p-bits typically require additional circuitry (transistors) to generate the desired tunable randomness in their implementations[17].

Here, we introduce perovskite nickelates as efficient probabilistic computing building blocks utilizing their intrinsic stochastic switching. The electrical properties of rare-earth nickelates are governed by strong electron-electron correlations[18]. For example, perovskite $NdNiO_3$ (NNO) are correlated metals at room temperature[19], however electron doping with hydrogen donors in NNO causes several orders of increase in the electrical resistivity. During hydrogenation, the protons ($H^+$) reside in the interstitial sites of $NiO_6$ octahedra, while the electrons from the hydrogens occupy Ni $e_g$ orbitals, opening up a large Mott gap and changing the electronic structure[20]. Owing to their small size of the order of $10^{-4}$ Å, protons can take on multiple metastable configurations with subtle changes to electronic structure[21]. By perturbing the hydrogen ions locally near Pd electrodes with electric fields, p-bit functionality is first demonstrated. We then introduce the algorithmic concept of synapse-sharing where all replicas of an original network are driven by the synaptic inputs obtained only from the main replica, greatly simplifying design and device operation while enabling a high degree of parallelism (**Figure 1(a)**). The compact combination of stochastic activation and synaptic summation obtained from simple perovskite nickelates devices is in stark contrast with other p-circuit implementations where the synapse operation is typically achieved by large external CMOS circuitry. We demonstrate feasibility in sampling as well as combinatorial optimization (**Figure 1(b)**). For a given hard optimization problem, there is no general method of knowing how many samples will be required. However, reducing the energy taken per sample and obtaining more probabilistic samples in compact hardware improve key metrics including energy-to-solution and time-to-solution. The framework we present in this paper helps with both these metrics: perovskite nickelate based p-bits reduce energy per random bit and area per random bit, with promising scalability in hardware. By activating multiple arrays of p-bits with the same synaptic input, the shared synapse algorithm leads to parallelism in the sampling throughput.

## Results and discussion

The stochastic switching behavior of our nickelate devices are summarized in **Figure 2**. NNO films (50 nm) were deposited on LaAlO$_3$ (LAO) substrates and structural characterization of pristine NNO films are shown in **Supplementary Figure 1**. After the hydrogen doping, the conductance of the nickelate films decreased by 4 orders of magnitude (**Supplementary Figure 1(d)**). Optical image of hydrogen doped NNO (H-NNO) p-bit device with schematic of electrical measurement setup is shown in **Figure 2(a)**. To investigate switching variation of the nickelate device, cycling tests and statistical analysis of H-NNO devices were performed, as shown in **Supplementary Figure 2** and **3**. By applying consecutive voltage pulses to the Pd electrode on H-NNO p-bits, abrupt changes in the resistance states of the device were observed (**Figure 2(b)**). The stochastic switching behavior of H-NNO p-bits in response to different pulse magnitude and width is presented in **Figure 2(c)**. As voltage pulse increases, a sigmoidal distribution of spiking probability of H-NNO p-bit device was obtained and can be tuned by the pulse width. We characterized our experiments with a probabilistic description of a two-state system separated by an energy barrier using a 1-D Fokker Planck Equation (FPE), an approach successfully utilized to describe bistable magnetic systems[22]. We assume that two states corresponding to high resistance state ($R_H$) and low resistance state ($R_L$) are separated by an energy barrier, that is consistent with experimental measurements. We also assumed that the switching voltage acts as a driving force that modulates the energy minima of the two-state system, preferentially minimizing the energy of one state with respect to the other. This situation is reminiscent of magnetic systems being driven by spin-polarized currents or magnetic fields (see **Methods** section for details of the FPE model). Using the same set of fitting parameters, the model was used predictively to describe different pulse width behavior, resulting in good quantitative agreement between the experimental results and the theoretical model. The logic configuration of H-NNO p-bits upon voltage pulses with different pulse magnitude are presented in **Figure 2(d)** and **Supplementary Figure 4**. As both pulse voltage magnitude and width increased, the population of the logic state 1 ($R_L$) became more dominant, indicating changes in hydrogen configuration in the NNO lattice. To demonstrate proof-of-concept use cases of our H-NNO p-bit for probabilistic computing, H-NNO p-bits were connected in various configurations and their probabilistic switching behavior were studied via experiment and simulations, as shown in **Supplementary Figure 6-15**. Details of simulation of connected H-NNO p-bits can be found in **Supplementary Figure 6** and **7**. Switching probability histograms of connected H-NNO p-bits demonstrated good agreement between experiment and simulated cases, as shown in **Supplementary Figure 11-13**.

In order to understand microscopic origins of the physical mechanism of stochastic switching behavior of H-NNO p-bits, we have performed reinforcement learning (RL) based first principles calculations. **Figure 3** shows the probability of finding a metastable configuration in a hydrogen-doped nickelate unit cell – the states are found to be distributed over an energy gap of ~0.8 eV. The metastable configurations were sorted with respect to their energies (left to right). The degree of non-equilibrium or metastability is shown by a corresponding heat map that depicts the configurational energy differences $\Delta E_i$ of all the metastable configurations with respect to the configuration which has the lowest energy. Our first principles calculations show that pristine NNO is metallic whereas the introduction of H within the NNO lattice opens up a band-gap. For a

1H:1NNO doping fraction, the probability of finding a given metastable H:NNO configuration (LHS y-axis) at temperature T=300 K is shown by the black curve and their corresponding gap ($E_g$) values are shown on the RHS y-axis. Representative metastable configurations for 1H:1NNO are presented in **Supplementary Figure 16**. We note a wide distribution of resistance states for the given stoichiometry and find that the resistance states are strongly correlated to the extent of metastability. In general, we find that near-equilibrium or low energy metastable 1H:1NNO configurations have a bandgap that varies from 0.2 to 0.8 eV. At 300 K, the most probable band gap value $<E_g>$, (calculated using $P_i \times E_{gi}$ i.e. probability of metastable state i multiplied with its band gap) is 0.52 eV as shown by the horizontal grey line. Configurations far away from energy minima tend to display a metallic behavior but are much less probable in the absence of an external stimuli. Most of the low energy configurations have semiconducting characteristic with subtle changes in the H locations in the NNO lattice and their band-gaps can be modulated with relatively low electric fields.

With the model for stochastic switching of the nickelate device, we show that H-NNO p-bits can be implemented in synapse-sharing algorithms (**Figure 4(a)**). In this algorithm, we construct a network in an N × N crossbar array with $N^2$ p-bits. These p-bits are split into N replicas (a column of the crossbar) where one replica is designated to be the main and the rest designated to be shared replicas. The key idea is to perform the multiply-accumulate (MAC) operation within the main replica, where a local field is computed for a given p-bit based on the state of the neighbors. All replicas of the same p-bit are updated with the same MAC input. The system involves a controller unit interacting with the crossbar array. Probabilistic circuits (p-circuits) employ two main equations, one for the stochastic activation and another for the synaptic operation. Typically, the stochastic activation is given by:

$$s_i = \Theta[-rand + \sigma(\beta I_i)] \quad \text{(Eq. 1)}$$

where $I_i$ is the dimensionless analog input, s is the binary state of a p-bit (either 0 or 1), $\beta$ is the dimensionless inverse temperature, and *rand* is a uniformly generated random number between 0 and 1. $\sigma$ is the sigmoid function ($\frac{1}{1+\exp(-x)}$) and $\Theta$ is the Heaviside step function. The synaptic equation corresponding to two-local interactions is given by:

$$I_i = \sum_j J_{ij} s_j + h_j \quad \text{(Eq. 2)}$$

where $J_{ij}$ are the weights between p-bits and $h_j$ are the bias terms for each p-bit. The main function of the controller array is to provide voltage pulses $V_{ij}$ and $V_i$ corresponding to the weights ($J_{ij}$) and bias ($h_j$) parameters of a chosen problem (**Figure 4(b)**). $R_H$ and $R_L$ of the H-NNO p-bits represent logic 0 and logic 1, respectively. Resistances ($R_j$) encode the stochastic state of the p-bits ($s_j$), whereas the voltage pulses applied from the controller encode the weight values ($J_{ij}$). The operation of a p-circuit consists of three main phases: deterministic dot product (read), stochastic switching (write) and read-out for bias correction (see **Supplementary Note 1.1**). Given a probabilistic network, the problem is introduced to the system by the weights ($J_{ij}$) mapped to a voltage set $V_{ij}$, and a bias ($h_i$) vector mapped to the voltage set $V_i$ (see details of mapping in the

**Methods** section). These voltages are chosen such that the system eventually samples from the corresponding Boltzmann distribution[23]. A digital controller is used to arrange $V_{ij}$, $V_i$ and $V_{write}$ voltages for the read ($V_{ij}$, $V_i$) and write phases ($V_{write}$). During the read phase, i$^{th}$ row of the weight matrix is chosen by the control unit (with some prescribed order), along with the i$^{th}$ bias voltage ($V_i$). The i$^{th}$ unit of the chosen weight array is excluded since the diagonals of the weight matrix is zero. As typically performed in crossbar arrays, the natural addition of currents according to Kirchhoff's Current Law (KCL) resulting in:

$$I_{MAC} = \sum_j \frac{V_{ij}}{Rj} + \frac{V_i}{R_{bias}} \quad \text{(Eq. 3)}$$

Given a row of p-bits (i$^{th}$ row) to be updated, this equation naturally maps to Eq. 2. After, the MAC current is turned into a voltage and digitized by an analog to digital converter, it goes back to the controller circuit (read phase complete). Next, a pulse voltage V$_{write}$ corresponding to I$_{MAC}$ is applied to i$^{th}$ row-line of the crossbar, updating the i$^{th}$ p-bits in all replicas (write phase complete). Subsequent p-bits in each replica are updated by consecutive read and write operations in this manner. In the shared synapse algorithm, N p-bits are updated per each write phase where each p-bits has access to its own uncorrelated randomness. This introduces natural parallelism to our algorithm, as we discuss next in solving hard combinatorial optimization problems. **Figure 4(c)** shows a neural network representation of the overall probabilistic network with the shared synapse algorithm. Discussion on the shared synapse algorithm is also provided in **Supplementary Note 1.2**.

We then demonstrated probabilistic Boolean gates utilizing H-NNO p-networks. In traditional digital design, Boolean gates are designed to be feedforward such that the input bits uniquely determine the output bits and the outputs are isolated from the inputs. This is an important (and desired) feature of digital gates enabling hierarchical design. By contrast, probabilistic gates can be designed to be bidirectional, enabling unique functionalities. For example, probabilistic AND gates can not only compute an output with respect to given inputs, but they can also be operated in the reverse mode where given an output, the circuit finds the input bits consistent with these outputs[24,25]. As we discuss below, this feature is useful in encoding optimization problems in probabilistic networks. In solving combinatorial optimization problems by probabilistic methods, typically an energy function is introduced:

$$-E = \sum_{<i,j>} J_{ij} s_i s_j + \sum_j h_j s_j \quad \text{(Eq. 4)}$$

Generally, an energy function is expressed heuristically[26] whose ground state corresponds to a problem of interest which is then sampled with high probability according to the Boltzmann distribution:

$$p = \frac{1}{Z} \exp(-E) \quad \text{(Eq. 5)}$$

Instead of expressing the energy functions heuristically, we follow the systematic composition of invertible Boolean gates[25,27]. This approach allows a sparse and systematic formulation of cost functions to express hard optimization problems. As a proof-of-principle, we focus on the problem of integer factorization and Boolean satisfiability.

**Figure 5(a-b)** shows the solution of the integer factorization problem by a probabilistic network in an 8-bit invertible multiplier[28]. This requires 52 p-bits in each replica. In an N × N crossbar layout, this corresponds to a total of 52 × 52 = 2704 p-bits in the system. For the main replica, a linear simulated annealing schedule is applied by changing $\beta$ (Eq. 1) from 0 to 1.5 linearly (see **Figure 5(a)** and **(c)** inset). Each sample corresponds to an updating event for a single p-bit in a replica, and in total, the system, including the shared replica takes $52 \times 5 \times 10^4$ samples. **Figure 5(a)** shows the energy as a heatmap for all replicas with respect to samples taken, when $\beta$ is increased linearly. We observe that at the end of an annealing schedule, the correct factors are found (**Figure 5(b)**). We see that the replica energies are also systematically converging to the ground energy but with different paths, indicating that the replicas explore useful alternatives in the rough energy landscape of the problem, a key advantage of the shared synapse algorithm. Next, we show an example Boolean satisfiability (3-SAT) problem, expressed in terms of invertible p-circuits (**Figure 5(c-d)**). Like integer factorization, Boolean satisfiability is a hard constraint satisfaction problem, widely used in many different applications ranging from planning in AI, software verification, and real-world optimization problems[29]. We choose an example instance, namely the UF920-91 from the UBC SAT library[30] with 20-literals and 91-clause as a representative benchmark. For this problem, we require 112 p-bits for each replica and therefore construct a 112 × 112 crossbar array with 12544 p-bits. We use a linear annealing schedule, changing $\beta$ from 0 to 2 in $10^4$ steps. Like the integer factorization, the energy of each replica is presented as a heatmap which shows how replicas take different paths but converge to the same, correct answer, despite being driven by the main replica in the shared synapse algorithm. Both results demonstrate the potential of probabilistic computing with inherently stochastic H-NNO p-bits as well as the use of additional probabilistic samples that can be obtained by the shared synapse method. Modifications to our algorithm to perform parallel tempering where replicas are driven at different temperatures by the same synaptic input from the main replica can lead to highly parallelized and hardware-aware sampling schemes.

Finally, benchmarks for random number generation using perovskite-based p-bits are presented for reference in **Supplementary Figure 21**. Random number generation (pseudo or true) is often a costly operation in digital CMOS circuits. CMOS implementations for pseudo-random number generators often require large circuits such as linear feedback shift registers which require thousands of transistors. More sophisticated pseudo random number generation with better random number quality, e.g., Xoshiro[31] is possible, but these introduce additional overhead in terms of area and energy needed per random bit. Therefore, to perform a fair and more stringent comparison, we focus on true random number generation with CMOS devices. Two representative implementations are with mixed-signal[32] (ring-oscillator jitter-based) and all-digital[33] (cross-coupled inverter based) circuits. The area and energy per random bits obtained from H-NNO p-bits can be substantially better than existing TRNG solutions.

## Conclusions

We have introduced nanoscale perovskite nickelate crystals as a p-bit platform for probabilistic computing. Electrical properties of individual and connected nanoscale devices have been studied via combination of experiments and modeling. Probabilistic networks simulated with experimentally measured stochastic switching behavior of the nickelate p-bits efficiently solved combination optimization problems such as integer factorization and Boolean satisfiability. The results suggest the potential of light-element doping of correlated perovskite crystals as an interesting platform for future computing technologies.

**Acknowledgements:** The analysis of the switching properties and related measurements were supported by Quantum Materials for Energy Efficient Neuromorphic Computing (Q-MEEN-C) an Energy Frontier Research Center (EFRC) funded by the U.S. Department of Energy (DOE), Office of Science, Basic Energy Sciences (BES), under Award DE-SC0019273. The fabrication and pulsed field measurements were supported by AFOSR FA9550-19-1-0351 and ARO W911NF-19-2-0237 respectively. K. Y. C. and K. S. acknowledge National Science Foundation support through CCF 2106260. Use of the Center for Nanoscale Materials, an Office of Science user facility, was supported by the U. S. Department of Energy, Office of Science, Office of Basic

Energy Sciences, under Contract No. DE-AC02-06CH11357. This material is based on work supported by the DOE, Office of Science, BES Data, Artificial Intelligence and Machine Learning at DOE Scientific User Facilities program (Digital Twins). This research used resources of the National Energy Research Scientific Computing Center, a DOE Office of Science User Facility supported by the Office of Science of the U.S. Department of Energy under Contract No. DE-AC02-05CH11231. We gratefully acknowledge the computing resources provided on Fusion and Blues, high performance computing clusters operated by the Laboratory Computing Resource Center (LCRC) at Argonne National Laboratory. This research used resources of the Advanced Photon Source, a U.S. Department of Energy (DOE) Office of Science user facility operated for the DOE Office of Science by Argonne National Laboratory under Contract No. DE-AC02-06CH11357.

**Competing interests:** The authors declare no competing interests.

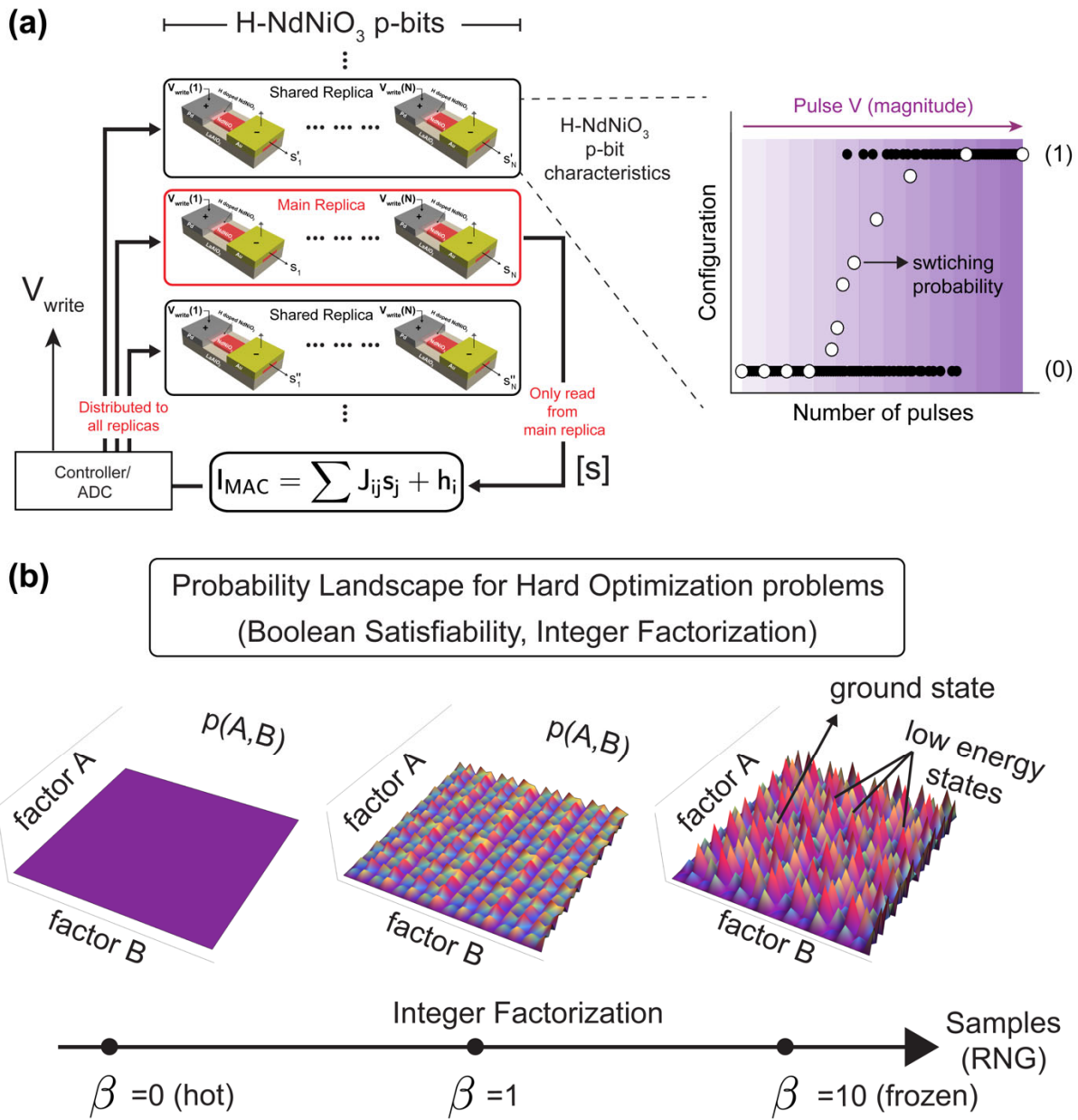

**Figure 1. Perovskite bits for probabilistic computing (a)** Probabilistic computer architecture with H-NNO p-bits utilizing a shared synapse algorithm. Main replica p-bits are used to compute a synapse to drive the full system with replicas. Each H-NNO acts as a p-bit, stochastically changing states from HRS (0) to LRS (1) by an applied pulse. Average switching probability was obtained at different pulse amplitudes. **(b)** Probability landscape for the integer factorization problem, representative of hard optimization problems. Algorithmic temperature of the system is decreased gradually to find the most favorable state among others for a defined problem. Roughness of the landscape indicates hardness, typical for hard optimization problems.

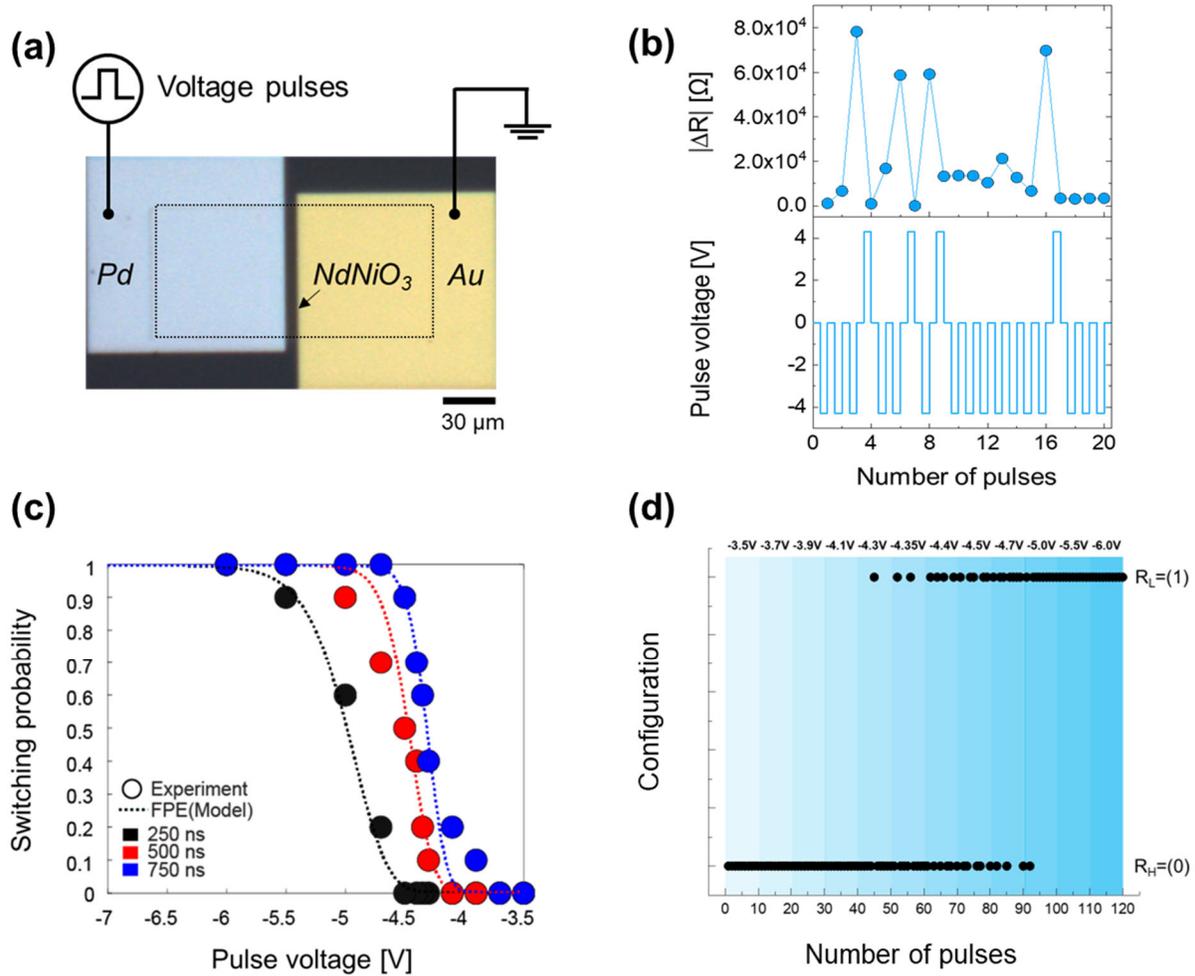

**Figure 2. Measured probabilistic switching behavior of H-NdNiO₃ p-bit device. (a)** Optical microscope image of H-NNO p-bit device and the electrical measurement scheme. Pd and Au electrodes were patterned on NNO films (dashed square). Voltage pulses were applied to Pd electrode and Au electrode was grounded. Scale bar, 30 μm. **(b)** Resistance changes of H-NNO p-bit device was monitored upon consecutive voltage pulses (top panel). Pulse voltage profile to the nickelate device as a function of number of pulses were presented (bottom panel). Negative voltage pulses (-4.5V, 500 ns) were applied to the nickelate device at high resistance state ($R_H$) until the resistance states were switched. After the switching, positive reset voltage (+4.5V, 500 ns) was applied to recover $R_H$ state. **(c)** Switching probability of H-NNO p-bit device as a function of applied pulse voltage with different pulse widths. The Fokker-Planck-Equation has been used to model the pulse width and pulse voltage dependence. **(d)** Logic configuration of H-NNO p-bit device, 0 and 1 denote $R_H$ and $R_L$, respectively. For a fixed pulse width of 500 ns, gradual increase in the population of the logic 1 state was observed as applied pulse voltage increased.

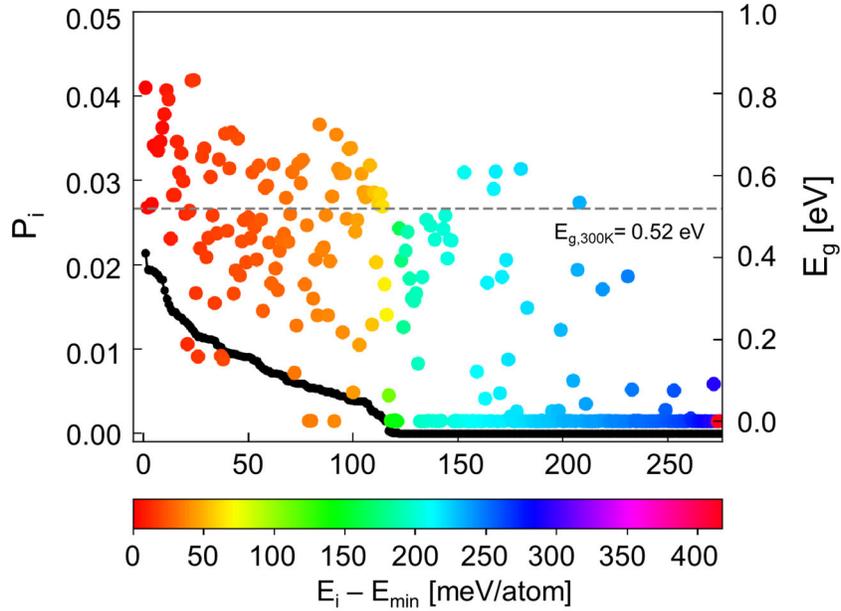

**Figure 3. Reinforcement learning-based first principles calculations to determine probability distribution plot for locating a metastable 1H:1NNO configuration at T= 300 K.** We sampled 275 different low energy metastable configurations which are shown in x-axis based on their energy ordering relative to the lowest energy 1H:1NNO configuration (from left to right with incresing energy in meV/atom as shown by the heatmap). Left y-axis shows the probability of finding metastable configuration i *i.e.* $P_i$ and RHS y-axis shows their corresponding band gap values calculated using a DFT+U framework. At T=300K, the most probable bandgap value for 1H:1NNO stoichiometry is 0.52 eV as shown by the horizontal grey line. The band-gap values vary widely and continuously over an 0.8 eV range. The application of an e-field can be used to perturb H within the NNO lattice – we note that subtle H displacements within the NNO lattice can allow us to access a wide range of semiconducting resistance states whereas higher fields can even allow for metallic states to be accessed (which typically has a low probability in the absence of any e-field).

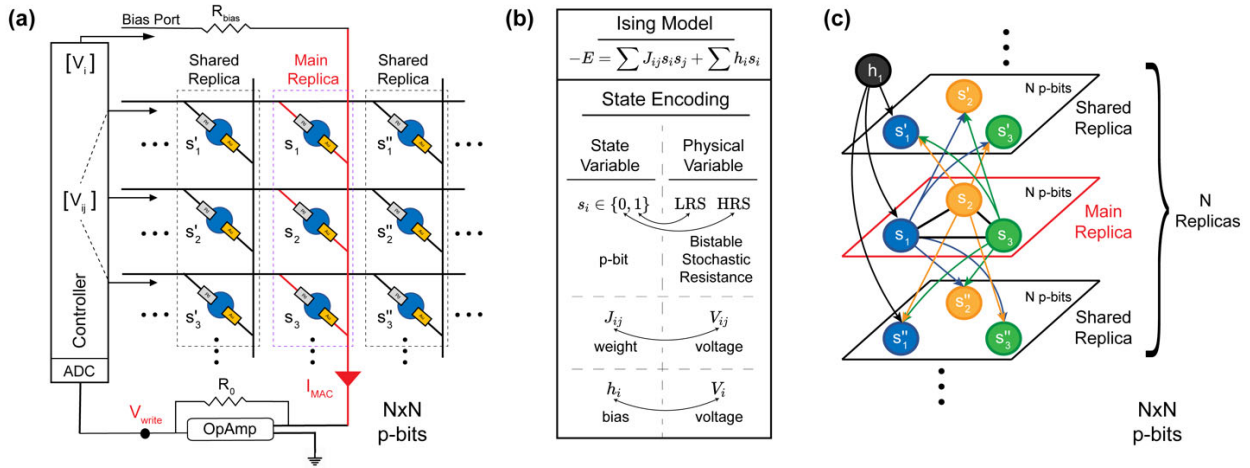

**Figure 4. Probabilistic computing architecture using resistive crossbar arrays, state encoding and the shared synapse algorithm.** **(a)** Physical structure of the crossbar array with N × N H-NNO p-bits. A multiply accumulate current ($I_{MAC}$) for a single p-bit is obtained over the main replica with N p-bits, converted to a voltage ($V_{write}$) and digitized by an ADC. Then, a full row of N p-bits (over N-replicas including the main) are updated with the shared ($V_{write}$) voltage. **(b)** Ising model and state encoding: low resistance and high resistance states (LRS, HRS) correspond to stochastic binary values, 1 and 0, respectively. Weights ($J_{ij}$) and biases ($h_i$) are described by voltages $V_{ij}$ and $V_i$ applied from the controller. **(c)** A neural network representation of the shared synapse algorithm: The inputs obtained from main replica p-bits drive neighboring p-bits in shared replicas in addition to the main replica. Biases are applied individually to each replica (only shown for $s_1$, for simplicity).

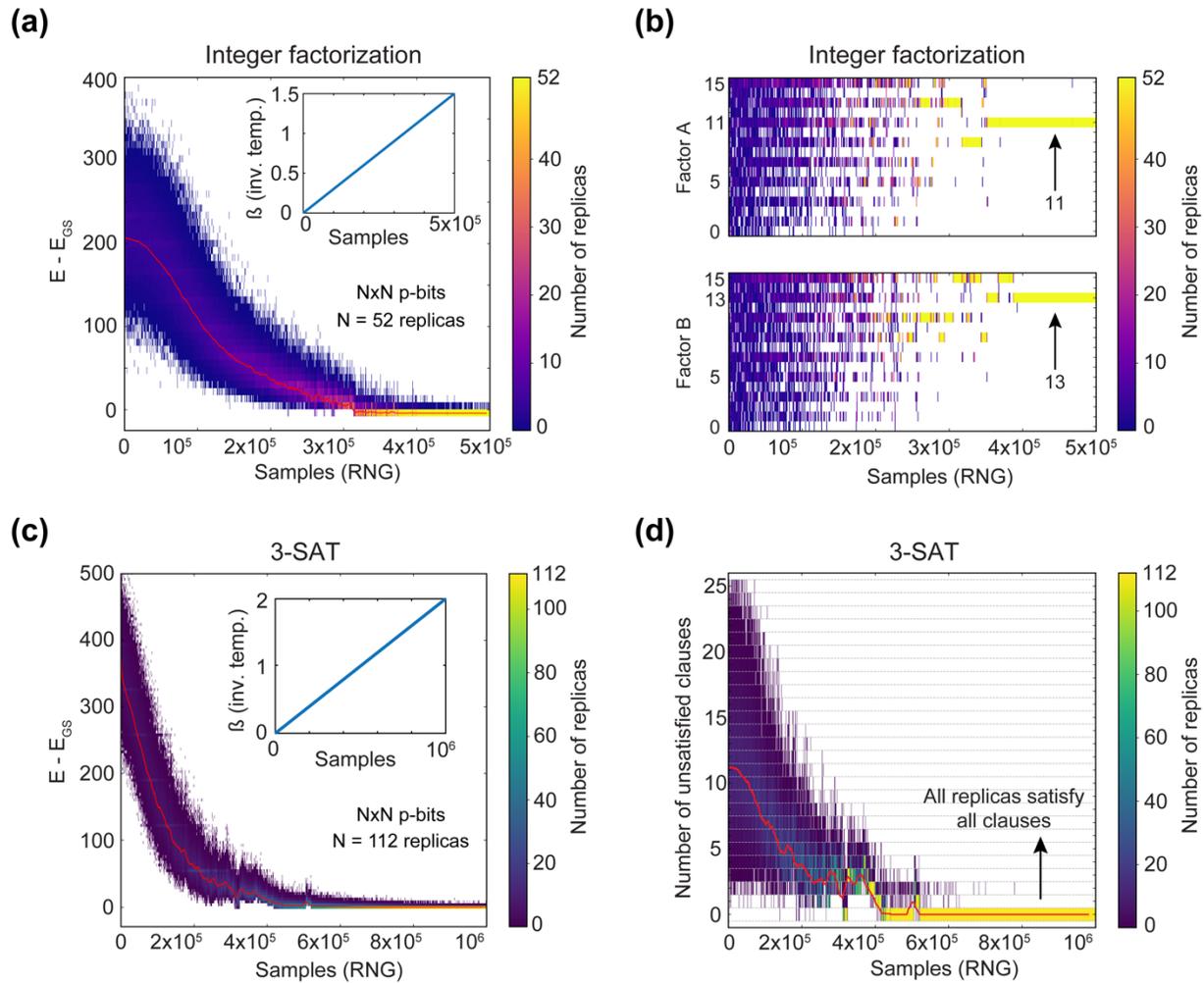

**Figure 5. Solving the Integer Factorization and Boolean Satisfiability (3-SAT) problems with the shared synapse algorithm (a)** Offset energy (where $E_{GS}$ is the ground state energy) as a function of samples size for an 8-bit integer factorization problem ($143 = 11 \times 13$). The simulation is performed on a $52 \times 52$ p-bit grid with 2704 p-bits with annealing over $5 \times 10^5$ samples. **(b)** Factors A and B when the product is fixed to 143 in an inverse multiplier p-circuit, measured over the entire range of samples. Heatmap of factors A and B measured as the system is annealed over $5 \times 10^5$ samples. **(c)** Offset energy as a function of sample size for the UF20-91 instance of the 3-SAT (Boolean Satisfiability) problem. (Inset) Shows linear annealing schedule of the inverse temperature, β. The energy is a heatmap over all replicas (N=112) where 111 shared replicas are driven by the main replica. Red line shows the average energy over all replicas. **(d)** Number of unsatisfied clauses in the 3-SAT instance, as a function of sample size showing a heatmap over all replicas. The replica heatmap shows how each replica individually converges to the right answer, though showing individual variations (Red line shows average replica energy). In both examples, despite sharing the synaptic input from the main replica, being endowed with uncorrelated random p-bits shared replicas show variations over the course of annealing, eventually converging to the same (correct) answer.